\documentclass[aps,twocolumn,showpacs]{revtex4-1}
\usepackage{graphicx}
\usepackage{amsmath}

\newcommand{\eps}{\varepsilon}
\newcommand{\bfr}{\boldsymbol{r}}	
\newcommand{\grad}{\boldsymbol{\nabla}}	
\newcommand{\del}{\partial}                     
\newcommand{\GeV}{{\rm GeV}}			
\newcommand{\MeV}{{\rm MeV}}			
\newcommand{\fm}{{\rm fm}}                      
\newcommand{\EoS}{equation of state} 

\begin{document}
\title{Spinodal density enhancements
in simulations of relativistic nuclear collisions}
\author{Jan Steinheimer} \email{jsfroschauer@lbl.gov}
\author{J{\o}rgen Randrup}

\affiliation{
Nuclear Science Division, Lawrence Berkeley National Laboratory,
Berkeley, California 94720, USA}

\date{October 9, 2012}

\begin{abstract}
We recently introduced a fluid-dynamical model for simulating relativistic nuclear collisions in the presence of a first-order phase transition and made explorative studies of head-on lead-lead collisions. We give here a more detailed account of this novel theoretical tool and carry out more exhaustive studies of the phase-separation dynamics. Extracting the density enhancement caused by the spinodal instabilities, the associated clump size distribution, and the resulting transverse flow velocity, we examine the sensitivity of these quantities to the strength of the gradient term that promotes the phase separation, to the details of the initial density fluctuations that form the seeds for the subsequent amplification, and to the equation of state.
\end{abstract}


\maketitle

\section{Introduction}
The aim of high-energy nuclear physics is to elucidate the properties 
of strongly interacting matter at extreme temperatures and densities.
A central subject in these investigations is the quark-gluon plasma (QGP),
a new phase of matter in which the quarks and gluons are deconfined.
Experiments at SPS, RHIC, and LHC have produced vast amounts of data 
indicating that such a new phase is indeed produced 
\cite{Afanasiev:2002mx,Gazdzicki:2004ef,Kumar:2011us,Adams:2005dq,%
Back:2004je,Arsene:2004fa,Adcox:2004mh,Aamodt:2010cz,Aamodt:2010jd,%
Aamodt:2010pa,Aamodt:2010pb}. 

Strongly interacting matter is expected to posses a rich phase structure.
In particular, 
compressed baryonic matter may exhibit a first-order phase transition 
that persists up to a certain critical temperature \cite{Stephanov:1998dy}. 
At vanishing baryon chemical potential, 
lattice QCD calculations yield a smooth transformation 
from confined to deconfined matter at a cross-over temperature 
of $T_\times\approx 150$ - $160$ MeV \cite{Borsanyi:2010cj,Bazavov:2010sb}. 
Unfortunately, the lattice calculations cannot be readily extended 
to finite chemical potential due to the so-called sign problem.
While several methods have been employed for circumventing this basic problem,
including Taylor expansion in $\mu/T$, 
reweighting, and imaginary chemical potential,
the results are not conclusive on the existence of a genuine first-order
phase transition at finite chemical potential
and the location of the associated critical point
 \cite{Fodor:2001pe,Fodor:2004nz,deForcrand:2008vr,Endrodi:2011gv}.

It is therefore necessary to rely on experiment for the determination 
of the phase structure of baryon-rich strongly interacting matter.
Experimental efforts are underway to search for evidence of 
a first-order phase transition and the associated critical end point 
\cite{RHIC-BES,CBM-book,NICA}.
For these endeavors to be successful, it is important to identify
observable effects that may serve as signals of the phase structure.
This is a challenging task because the colliding system is 
relatively small, non-uniform, far from global equilibrium,
and rapidly evolving, 
features that obscure the connection between experimental observables and
the idealized uniform equilibrium matter described by the equation of state.
Therefore, to understand how the presence of a phase transition 
may manifest itself in the experimental observables,
it is necessary to carry out dynamical simulations of the collisions
with suitable transport models.

Many numerical simulations of high-energy nuclear collisions
have employed ideal or viscous fluid dynamics which has
the important advantage that the equation of state (EoS) appears explicitly.
By contrast, in most microscopic transport models
the EoS is often unknown or very cumbersome to determine. 
The focus up to now has mainly been on bulk observables 
and their dependence on a softening of the EoS. 
For this purpose, the instabilities associated with a first-order
phase transition were usually removed by means of a Maxwell construction,
thereby ensuring that bulk matter remains mechanically stable
throughout the expansion.

However, when a first-order phase transition exists,
a low-density confined phase (a hadronic resonance gas)
may coexist thermodynamically with a high-density deconfined phase
(a baryon-rich quark-gluon plasma) and, consequently,
bulk matter prepared at intermediate densities would be unstable
and seek to separate into the two coexisting phases.
Therefore, in a nuclear collision,
when the dynamical evolution drives the bulk density
into the phase coexistence region, the instabilities will be triggered.
In particular, the spinodal instabilities
\cite{PhysRep,Randrup:2003mu,Sasaki:2007db,Randrup:2009gp}
may generate a non-equilibrium evolution that in turn
may generate observable fluctuations in the baryon density
\cite{RandrupAPH22,KochPRC72,Mishustin:1998eq,Bower:2001fq}
and the chiral order parameter \cite{Paech:2003fe,NahrgangPRC84}.
Furthermore, nucleation and bubble formation 
may also contribute towards the phase separation process.

In order to ascertain the degree to which these mechanisms may 
manifest themselves in actual nuclear collisions,
we have performed numerical simulations with finite-density fluid dynamics,
using a previously developed two-phase equation of state
and incorporating a gradient term in the local pressure
\cite{SteinheimerPRL109}.
This latter refinement emulates the finite-range effects 
that are essential for a proper description of the phase transition physics
\cite{PhysRep,Randrup:2003mu,Randrup:2009gp,Randrup:2010ax}.
In particular, the gradient term ensures that two coexisting bulk phases
will develop a diffuse interface 
and acquire an associated temperature-dependent tension.
Furthermore, of key importance to the present study,
the gradient term also causes the dispersion relation for the collective modes
in the unstable phase region to exhibit a maximum,
as is a characteristic feature of spinodal decomposition \cite{PhysRep}.
Thus we employ a transport model 
that has an explicitly known two-phase equation of state 
and that treats the associated physical instabilities
in a numerically reliable manner.

A first application of this novel tool was recently made for 
head-on collisions of lead nuclei at various energies \cite{SteinheimerPRL109}.
For a certain range of collision energies, several GeV per nucleon,
the bulk of the system will reside within the spinodal region 
of the phase diagram for a sufficient time to allow the associated 
instabilities to significantly enhance the initial density irregularities.
This characteristic phenomenon was exploited previously 
to obtain experimental evidence for the nuclear liquid-gas phase transition
\cite{BorderiePRL86,PhysRep}
and we are exploring the prospects for it being useful
for signaling the confinement phase transition.

\section{The Equation of State}

\begin{figure}[t]	
\includegraphics[width=0.48\textwidth]{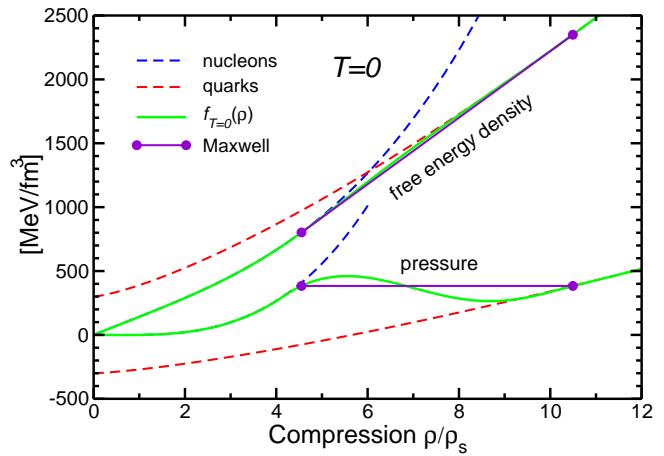}
\caption{[Color online] The construction of the \EoS\ is illustrated for $T=0$.
{\em Top part:} The free energy densities for either an interacting nucleon gas
or quarks in a bag are joined by a spline, yielding a free energy density
$f_{T=0}(\rho)$ (solid curve) having a concave anomaly,
as is characteristic of a first-order phase transition.
The single-phase partner $f_T^M(\rho)$ results when $f_T(\rho)$ 
is replaced by the common tangent in the coexistence density region.
{\em Bottom part:} The corresponding plot for the pressure $p_T(\rho)$,
which exhibits an undulation through the coexistence region where 
its Maxwell partner $p_T^M(\rho)$ remains equal to the coexistence pressure.
}\label{spline}
\end{figure}

In order to obtain a suitable equation of state, 
we employ the method developed in Ref.\ \cite{Randrup:2009gp}.
Thus we work (at first) in the canonical framework and, for a given $T$,
we obtain the free energy density $f_T(\rho)$ in the phase coexistence region 
by performing a suitable spline between two idealized systems
(either a gas of pions and interacting nucleons 
or a bag of gluons and quarks) held at that temperature.
In Ref.\ \cite{Randrup:2009gp} the focus was restricted to
 subcritical temperatures, $T<T_{\rm crit}$,
so for each $T$ the spline points were adjusted
so the resulting $f_T(\rho)$ would exhibit a concave anomaly,
{\em i.e.}\ there would be two densities, $\rho_1(T)$ and $\rho_2(T)$,
for which the tangent of $f_T(\rho)$ would be common.
This ensures phase coexistence, {\em i.e.}\ the chemical potentials match,
$\mu_T(\rho_1)=\mu_T(\rho_2)$,
because $\mu_T(\rho)=\del_\rho f_T(\rho)$, 
and so do the pressures, $p_T(\rho_1)=p_T(\rho_2)$,
because $p_T(\rho)=\mu_T(\rho)\rho-f_T(\rho)$.
Ref.\ \cite{SteinheimerPRL109} extended the \EoS\ to $T>T_{\rm crit}$
by using splines that are convex,
as is characteristic of single-phase systems.
After having thus constructed $f_T(\rho)$
for a sufficient range of $T$ and $\rho$,
we may obtain the pressure, as well as the energy density
$\eps_T(\rho)=f_T(\rho)-T\partial_T f_T(\rho)$, by suitable interpolation 
and then tabulate the \EoS, $p_0(\eps,\rho)$, 
on a convenient Cartesian lattice.
The procedure is illustrated in Fig.\ \ref{spline}.

\begin{figure}[t]	
\includegraphics[width=0.5\textwidth]{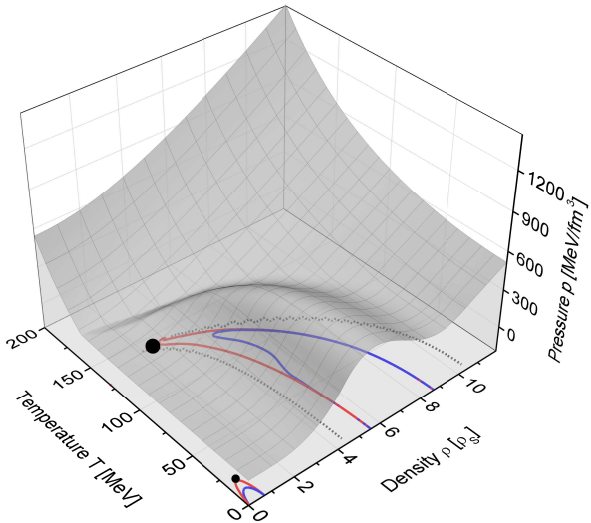}
\caption{[Color online] 
The canonical equation of state $p_T(\rho)$ constructed by a spline procedure, 
as described in the text.
The phase-coexistence boundaries (dotted lines) 
join at the critical point (solid dot).
Also shown are the isothermal spinodal boundary where $v_T\!=\!0$ (red)
and the isentropic spinodal boundary where $v_s\!=\!0$ (blue). 
The familiar nuclear liquid-drop phase transition occurs at
sub-saturation densities, while the confinement transition
occurs at densities well above the saturation value $\rho_s$.
}\label{f:1}
\end{figure}

Figure \ref{f:1} shows a contour plot of the resulting \EoS\ 
in the canonical representation, $p_T(\rho)$. 
The lower and upper boundaries of the phase coexistence region,
$\rho_1(T)$ and $\rho_2(T)$, respectively, are shown;
they come together at the critical point $(\rho_{\rm crit},T_{\rm crit})$.
Uniform matter is thermodynamically unstable inside this region,
but still mechanically stable near the phase coexistence boundaries.
But inside the spinodal boundary, along which the speed of sound vanishes,
uniform matter is both thermodynamically and mechanically unstable
so, consequently, arbitrarily small density irregularities are amplified
at a scale-dependent rate $\gamma_k(\rho,T)$.

It is important to recognize that the introduction of the gradient term
ensures that our model describes the instabilities 
not only in the unstable spinodal region,
but also those in the surrounding metastable region 
in which finite seeds are required for amplification to occur 
(yielding nucleation or bubble formation \cite{Mishustin:1998eq}).
Thus density irregularities may be amplified by the metastable region as well,
and any clumping generated inside the spinodal region may be further enhanced
as the system expands through the nucleation region.

When the dynamical evolution is governed by ideal fluid dynamics (see below)
the spinodal boundary is characterized by the vanishing
of the isentropic sound speed, $v_s=0$,
where $v_s^2\equiv(\rho/h)\left(\del p/\del\rho\right)_{s/\rho}$,
with $s$ being the entropy density and $h\!=\!p+\eps$ the enthalpy density.
For dissipative evolutions the finite heat conductivity causes the
instability region to widen and the boundary is then characterized by
the vanishing of the isothermal sound speed, $v_T=0$,
where $v_T^2\equiv(\rho/h)\left(\del p/\del\rho\right)_T\leq v_s^2$.
These spinodal boundaries are also indicated on Fig.\ \ref{f:1}.

\subsection{Single-phase equation of state}

In order to ascertain the dynamical effects 
of the first-order phase transition,
we construct a one-phase partner \EoS\ that resembles
the two-phase \EoS\ as much as possible.
Therefore no changes are made outside the phase coexistence region,
but for each subcritical temperature $T<T_{\rm crit}$
the two-phase free energy density 
$f_T(\rho)$ is replaced by the value obtained by performing a Maxwell 
construction through the coexistence density region,
{\em i.e.}\ for $\rho_1(T)<\rho<\rho_2(T)$, namely
\begin{equation}
f_T^M(\rho)=f_T(\rho_i)+(\rho-\rho_i)\mu(\rho_i)<f_T(\rho)\ .
\end{equation}
We note this construction lowers the free energy density.
This procedure is also illustrated in Fig.\ \ref{spline}.

Thus, at a given subcritical temperature $T$,
the chemical potential remains constant when the density 
is increased from $\rho_1(T)$ to $\rho_2(T)$,
because $\mu_T$ is the slope of $f^M_T(\rho)$ which is linear.
It then follows that also the pressure $p=\mu\rho-f$  remains constant
because it is linear in $\rho$ and matches at $\rho=\rho_i$,
$f^M_T(\rho_i)=f_T(\rho_i)$ for $i=1,2$.

\section{Fluid Dynamical Clumping}

For our present investigation, we describe the evolution of the
colliding system by ideal fluid dynamics,
because dissipative effects are not expected to play a
decisive role for the spinodal clumping \cite{Randrup:2010ax}:
Although the inclusion of viscosity generally tends to slow the growth,
the dissipative mechanisms responsible for the viscous effects
also lead to heat conduction which has the opposite effect
and also enlarges the unstable region
(from the isentropic to the isothermal boundary).

The basic equation of motion in ideal fluid dynamics
expresses four-momentum conservation, $\partial_\mu T^\mu=0$,
where the stress tensor is given by
\begin{equation}\label{T}
T^{\mu\nu}(x) = [p(x)+\eps(x)]u^\mu(x)u^\nu(x)-p(x)g^{\mu\nu}\ ,
\end{equation}
where $u^\mu(x)$ is the four-velocity of the fluid.
When taking account of the baryon current density,
\begin{equation}
	 N^\mu(x)=\rho(x)u^\mu(x)\ ,
\end{equation}
the basic equation of motion 
is supplemented  by the continuity equation, $\partial_\mu N^\mu=0$.
These equations of motion are solved by means of
the code SHASTA \cite{Rischke:1995ir}
in which the propagation in the three spatial dimensions
is carried out consecutively.

As mentioned above,
a proper description of spinodal decomposition requires that
finite-range effects be incorporated
\cite{PhysRep,Randrup:2003mu}.
Therefore, following Refs.\ \cite{Randrup:2009gp,Randrup:2010ax},
we write the local pressure as 
\begin{equation}\label{p}
p(\bfr)	=p_0(\eps(\bfr),\rho(\bfr))
-a^2\eps_s{\rho(\bfr)\over\rho_s}\grad^2{\rho(\bfr)\over\rho_s}\ ,
\end{equation}
where we recall that $p_0(\eps,\rho)$ is the \EoS,
the pressure in uniform matter characterized by $\eps$ and $\rho$.
With $\rho_s=0.153/\fm^3$ being the nuclear saturation density
and $\eps_s\approx m_N\rho_s$ the associated energy density,
the gradient term is normalized such that its strength
is conveniently governed by the length parameter $a$.

\subsection{Interface tension}

As for the pressure (see Eq.\ (\ref{p}) above), the free energy density 
is also modified by the presence of the gradient term.
At a given global temperature $T$, its local value is always increased,
\begin{equation}\label{f}
f(\bfr)= f_T(\rho(\bfr))
	+\mbox{$1\over2$}a^2\eps_s(\grad{\rho(\bfr)\over\rho_s})^2\ ,
\end{equation}
where $f_T(\rho)$ is the free energy density in uniform matter
at the specified density $\rho$ and temperature $T$.
Thus sharp changes in the density are costly and, consequently,
if two coexisting phases of bulk matter are brought into physical contact
a diffuse interface develops between them.
This feature is obviously of key importance for determining
the geometric properties of mixed-phase configurations,
such as the preferred size of droplets or bubbles.
Generally, the equilibrium density can be determined by minimizing 
the total free energy, subject to conservation of the total baryon number,
\begin{equation}
\delta\int[f(\bfr)-\mu\rho(\bfr)]d\bfr\ \doteq\ 0\ ,
\end{equation}
where $\mu$ is the coexistence value of the chemical potential.

The crucial quantity is the tension associated with the interphase
between the two coexisting bulk regions.
This quantity is most easily determined by considering a planar interface,
in which case the above variational condition,
after insertion of the form in (\ref{f}), yields \cite{Randrup:2009gp}
\begin{equation}\label{rhox}
a^2{\eps_s\over\rho_s^2}{\del^2\over\del x^2}\rho(x) =\mu_T(\rho(x))-\mu
= \left[{\del\over\del\rho}\Delta f(\rho)\right]_{\rho=\rho(x)} ,
\end{equation}
where the ``depth'' $x$ is measured in the direction normal to the interface
and $\mu_T(\rho)$ denotes the chemical potential in uniform matter
at the specified density $\rho$ and temperature $T$.
Furthermore, in the last expression we have used the relation 
$\mu_T=\del_\rho f_T$ to introduce the quantity
$\Delta f(\rho)\equiv f_T(\rho)-f_T^M(\rho)$,
the difference between the actual free energy density at the specified density
and that associated with the corresponding Maxwell construction;
it is negative for the relevant densities
(those between the two coexistence densities).

The solution of the above differential equation (\ref{rhox}) yields
the diffuse equilibrium density profile $\rho(x)$
which changes smoothly from one coexistence value to the other
as $x$ moves through the interface region.
However, due to the simplicity of the gradient approximation,
the associated interface tension can readily be determined from the \EoS\ 
alone, without explicit knowledge of $\rho(x)$ \cite{Randrup:2009gp}, namely
\begin{equation}
\sigma_T\ 
=\ a\int_{c_1(T)}^{c_2(T)}\left[2\eps_s\Delta f_T(c)\right]^{1/2}dc\ ,
\end{equation}
where $c\equiv\rho/\rho_s$ denotes the degree of compression.
At the specified sub-critical temperature $T$,
the integral extends over compressions in the corresponding coexistence region.
The interface tension $\sigma_T$ decreases steadily as $T$ is raised
and finally vanishes at $T_{\rm crit}$.
It is therefore convenient to characterize it by its maximum value,
$\sigma_{T=0}$.

Generally the tension between two coexisting phases plays a key role
in determining the characteristic spatial scale of the phase mixture that 
results when matter prepared in the unstable spinodal region phase separates.
Unfortunately, the strength of the tension between nuclear and quark matter
is not well understood
and the literature contains a large spread of suggested values.
For our present studies we have adjusted the range $a$ 
so that $\sigma_0\approx10\,\MeV/\fm^2$, which is obtained for $a=0.033\,\fm$.
(For comparison the surface tension of nuclear matter is
$\approx$$1\,\MeV/\fm^2$.)
Though this value lies in the lower part of the range,
it was preferred in a related study \cite{Mishustin:1998eq} and agrees with the value of the surface tension found for the Polyakov-Quark-Meson model \cite{Mintz:2012mz}
(see also the discussions 
in Refs.\ \cite{HeiselbergPRL70,VoskresenskyNPA723}, for example).

\subsection{Spinodal amplification}

As mentioned above, uniform matter inside the spinodal region (where $v_s^2<0$)
is mechanically unstable and density ripples of wave number $k$
will be amplified at a rate $\gamma_k(\rho,\eps)$.
Its value depends on the strength of the gradient term
and is given by the dispersion relation,
\begin{equation}
\gamma_k = 
\left[|v_s|^2k^2 -a^2{\eps_s\over h}{\rho^2\over\rho_s^2}k^4\right]^{1\over2} .
\end{equation}
The first term reflects the familiar linear rise
obtained for standard ideal fluid dynamics, $\gamma_k=|v_s|k$,
while the second term arises from the gradient term
which introduces a penalty for the development
of short-range undulations.
As a consequence of these two opposing trends,
the resulting growth rate $\gamma_k$ exhibits a maximum
at the favored length scale $\lambda_{\rm opt}=2\pi/k_{\rm opt}$.

The spinodal growth rates can be extracted by following the time evolution 
of small harmonic perturbations of uniform matter.
Thus, imposing periodic boundary conditions,
we consider initial systems of the form
\begin{equation}\label{init}
\rho(\bfr)=\bar{\rho}+\delta\rho(0)\sin(kx)\ ,\
\eps(\bfr)=\bar{\eps}+\delta\eps(0)\sin(kx)\ ,
\end{equation}
where $(\bar{\rho},\bar{\eps})$ lies inside the spinodal phase region
and the amplitudes $\delta\rho(0)$ and $\delta\eps(0)$ are suitably small.
Because the frequency is purely imaginary, $\omega_k=\pm i\gamma_k$,
the early time evolution of the amplitudes will consist of
growing and decaying exponentials having equal weights
(because the initial state (\ref{init})
is prepared without any flow) \cite{Lalime},
\begin{equation}\label{t}
\delta\rho(t)\approx\delta\rho(0)\cosh(\gamma_k t)\ ,\
\delta\eps(t)\approx\delta\eps(0)\cosh(\gamma_k t)\ ,
\end{equation}
and it is then straightforward to extract the rate $\gamma_k$
from the calculated amplitude growth.

\begin{figure}[t]	
\includegraphics[width=0.5\textwidth]{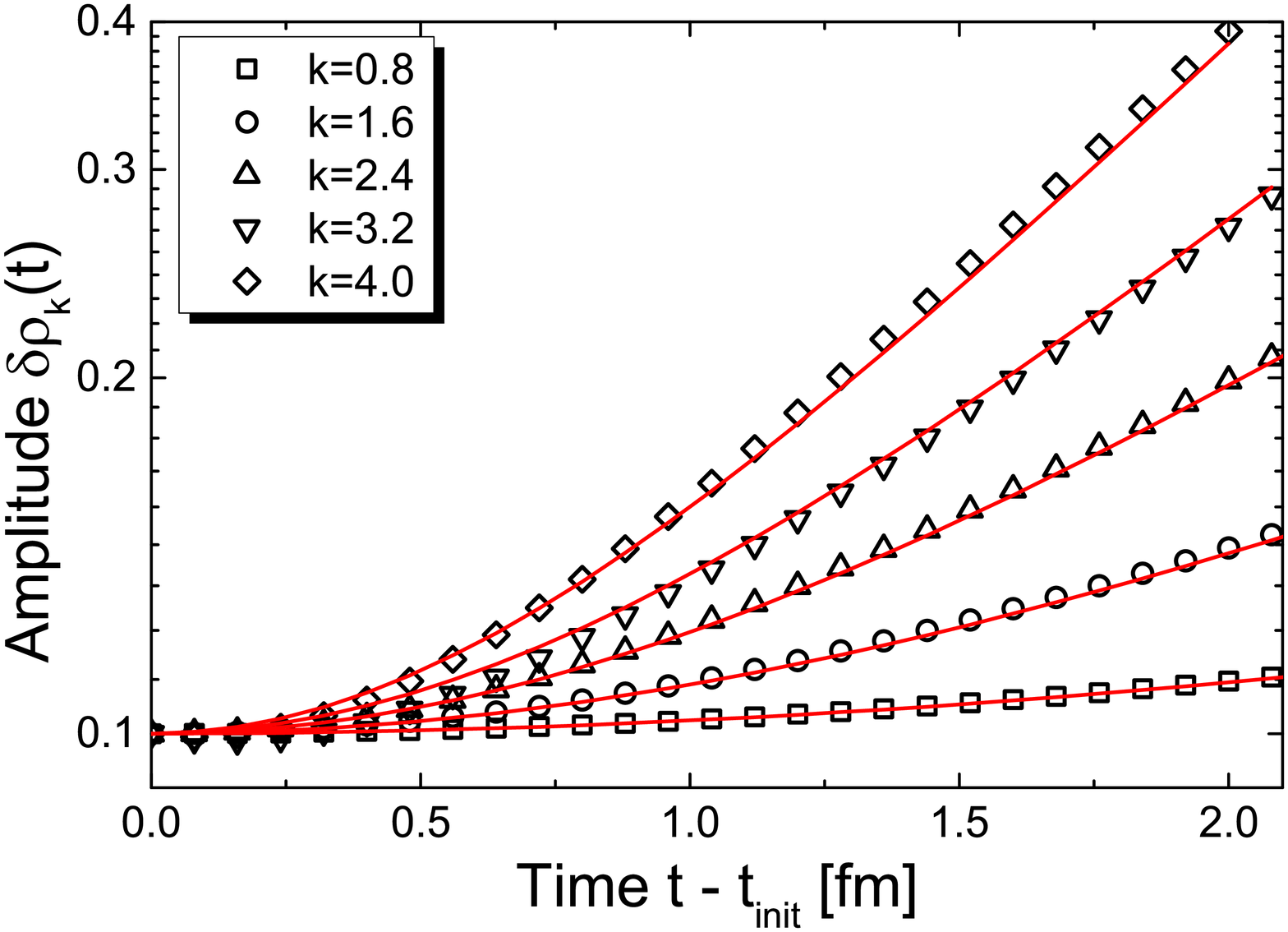}	
\caption{[Color online] The growth of harmonic density undulations
inside the spinodal region as obtained with standard ideal fluid dynamics,
shown by the symbols for various values of the wave number $k$,
together with the resulting fits to the expected analytical form (\ref{t}),
shown by the continuous curves.
}\label{fit}
\end{figure}		

This is illustrated in Fig.\ \ref{fit} for the phase point
$(\bar{\rho},\bar{\eps})=(6\rho_s,10\eps_s)$,
which lies well inside the spinodal region,
and using $(\delta\rho(0),\delta\eps(0))=(0.1\rho_s,0.2\eps_s)$.
The subsequent time evolution is obtained with ideal fluid-dynamics 
(without the gradient term for this illustration)
and the Fourier components of the density are extracted.
The resulting time-dependent amplitudes $\delta\rho_k(t)$
are then fitted with the analytical form (\ref{t}).
As the figure brings out,
the expected form is indeed produced,
indicating that the numerical propagation of the unstable system is reliable
and that the growth rates $\gamma_k$ can be extracted with confidence.

\begin{figure}[t]	
\includegraphics[width=0.5\textwidth]{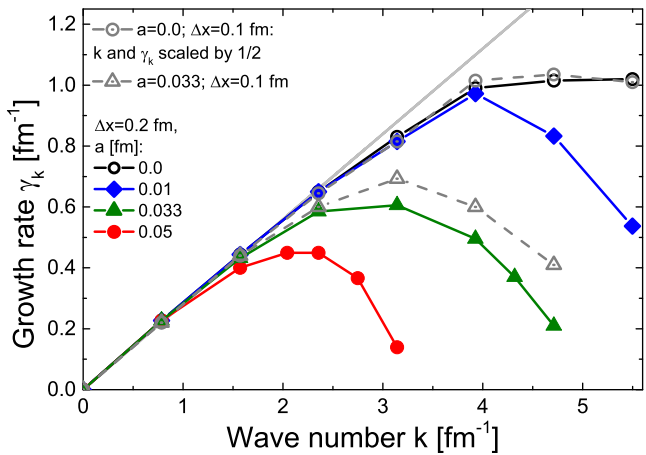}	
\caption{[Color online] Spinodal dispersion relation $\gamma_k$,
the growth rate as a function of the wave number.
The straight line is the exact result for gradient-free ideal fluid dynamics,
whereas the two top curves are the corresponding numerical results
obtained with the standard grid size $\Delta x=0.2\,$fm (solid)
or with $\Delta x=0.1\,$fm (dashed).
The four lowest curves were obtained with finite
values of the range $a$ as indicated.
}\label{f:gamma}
\end{figure}

Figure \ref{f:gamma} shows dispersion relations extracted in this manner
for various scenarios.
Ideal fluid dynamics without a gradient term yields the indicated straight 
line, $\gamma_k=|v_s|k$, and this behavior is indeed approached by the 
numerical results when the employed spatial grid size $\Delta x$ is reduced:
Our standard value of $\Delta x=0.2\,\fm$ yields the top solid curve,
while using of half that value, $\Delta x=0.1\,\fm$,
simply scales up the resulting curve $\gamma_k$ by a factor of two,
{\em i.e.}\ $\gamma_{2k}[\mbox{$1\over2$}\Delta x]=2\gamma_k[\Delta x]$.
The deviation of the numerical $\gamma_k$ from the analytical behavior 
$|v_s|k$ is due to the numerical procedure in the 
SHASTA code \cite{Rischke:1995ir}.

The other curves on Fig.\ \ref{f:gamma} have been calculated with
the gradient term included, using the indicated values of the range $a$.
They all lead to curves that exhibit a maximum growth rate
and we see that using $\Delta x=0.2\,\fm$ ensures that the finite-$a$
results are physically reasonable for range values down to $a=0.01\,\fm$.
Our studies suggest that while the gradient-free calculations 
with $\Delta x=0.2\,\fm$ are reliable only up to $k\lesssim4/\fm$ 
(beyond which the growth rate levels off towards a constant),
the use of a finite range, $a>0$,
extends validity to significantly finer scales
due to the suppression of such irregularities.
Furthermore, for ranges above $\approx$0.01\,fm
the growth rates are rather insensitive to the employed grid size $\Delta x$,
as is illustrated for our standard range of $a=0.033\,\fm$
(which yields a surface tension of $\sigma_0$$\approx$$10\,\MeV/\fm^2$,
as discussed above).

In actual systems there is some degree of physical dissipation
which gives rise to both viscosity and heat conduction.
To leading order, the viscosity reduces the growth rate by
$\approx\mbox{$1\over2$}[\mbox{$4\over3$}\eta+\zeta]k^2/h$,
where $\eta$ and $\zeta$ are the shear and bulk  viscosity coefficients,
respectively.
But the heat conductivity generally increases the growth rate
(in particular, it extends the boundary of the mechanically unstable region
from the isentropic to the isothermal spinodal).

\section{Nuclear collisions}

We have shown above that our model produces 
both meaningful interface properties,
including the temperature-dependent strength of the associated tension,
and reasonable spinodal growth rates,
including the emergence of an optimal phase separation scale.
The model is therefore suitable for addressing dynamical scenarios
involving phase-transition instabilities.

For a first study, we have considered central collisions of lead nuclei
bombarded onto a stationary lead target
at various kinetic energies, $E_{\rm lab}$.
For each energy,
an ensemble of several hundred separate evolutions are generated,
each one starting from a different initial configuration
generated by (the cascade mode of) the UrQMD model
\cite{Bass:1998ca,Bleicher:1999xi,Petersen:2008dd}
which treats the non-equilibrium dynamics 
during the very early stage of the collision. 
The switch from UrQMD to fluid dynamics is made at the time
\begin{equation}
	t_{\rm init} = {2R}/{\sqrt{\gamma_{C.M.}^2 -1}}\ ,
\end{equation}
where $\gamma_{C.M.}$ is the center-of-mass frame Lorentz factor 
and $R$ is the radius of the nucleus.
At this time the two Lorentz-contracted nuclei 
have just passed through each other
and all initial hard collisions have occurred. 
Each hadron is then represented by a gaussian of width $\sigma_{\rm init}=1$~fm
and the needed fluid-dynamical quantities,
$\rho(\bfr)$, $\eps(\bfr)$, $\boldsymbol{u}(\bfr)$, can then be extracted
and tabulated on a desirable Cartesian spatial lattice 
\cite{Steinheimer:2007iy}.
In this way, the effects of stopping as well as local event-by-event 
fluctuations are automatically included 
in the ensemble of initial fluid-dynamic states.

\subsection{Density clumping}

The amplification of spatial irregularities presents
an important effect of the presence of a first-order phase transition
\cite{Randrup:2003mu,Randrup:2009gp,Randrup:2010ax,PhysRep}.
In the present scenarios, spatial irregularities are present already in the initial state,
whereas the fluid-dynamical propagation does not generate any
spontaneous fluctuations in the course of the evolution
(such fluctuations are generally produced at finite temperatures 
\cite{Kapusta:2011gt}
but this refinement has not yet been incorporated into the
fluid-dynamical transport treatments of nuclear collisions).

For any given event, and at any given time, 
parts of the system may lie within the unstable or metastable region,
and local density irregularities may then become amplified,
whereas the rest of the matter is situated in a stable phase region
where irregularities tend to erode.
In order to ascertain the effect of those instabilities,
we also carry out corresponding simulations with the 
one-phase Maxwell partner \EoS\ which contains no instabilities
but is otherwise identical.

A convenient quantitative measure of the 
resulting degree of ``clumping'' in the system is provided by
the moments of the baryon density density $\rho(\boldsymbol{r})$,
\begin{equation}\label{moments}
\langle\rho^N\rangle\	\equiv\		{1\over A^N}
\int\rho(\boldsymbol{r})^{N}\rho(\boldsymbol{r})\,d^3\boldsymbol{r}\ ,
\end{equation}
where $A=\int\!\rho(\boldsymbol{r})d^3\boldsymbol{r}=\langle\rho^0\rangle$ 
is the total (net) baryon number.
The corresponding normalized moments,
$\langle\rho^N\rangle/\langle\rho\rangle^N$, 
are dimensionless and increase with the order $N$,
for a given density distribution $\rho(\bfr)$;
the normalized moment for $N=1$ is unity.

\begin{figure}[t]	
\includegraphics[width=0.5\textwidth]{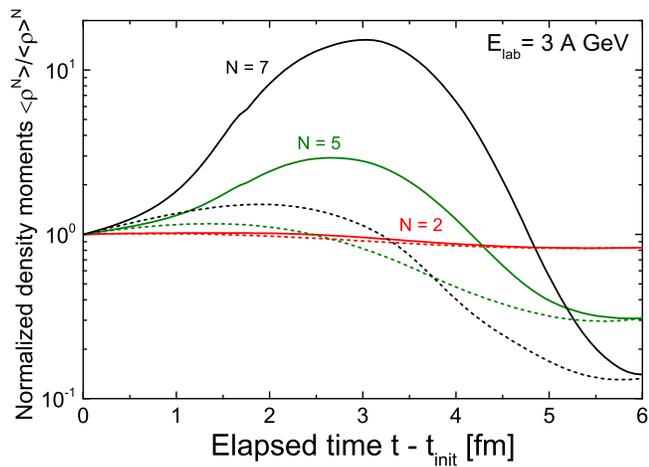}
\caption{[Color online] 
Mean time evolution of the amplification of normalized density moments
obtained for $E_{\rm lab}=3A\,\GeV$ with $a=0.033\,\fm$
(for which $\sigma_0\!\approx\!10\,\MeV/\fm^2$),
using either the two-phase equation of state (solid)
or its one-phase partner (dashed).
}\label{f:3}
\end{figure}		

The time evolution of the amplification of the normalized density moments
is illustrated in Fig.\ \ref{f:3} for $N=2,\dots,7$,
as obtained with both equations of state.
The two sets of results differ strikingly:
Whereas the one-phase \EoS\ hardly produces any amplification at all,
the one having a phase transition leads to significant enhancements,
amounting to well over an order of magnitude for $N=7$.
This clearly demonstrates that the first-order phase transition
may have a qualitative influence on the dynamical evolution of the density.
Due to the variations in the initial conditions,
the phase regions explored differ from one collision to the other.
As a result, some of the evolutions experience amplifications
that are significantly larger than the ensemble average
(by up to a factor of five or so),
whereas more evolutions are affected considerably less.

Figure \ref{f:3} depicts what would happen
if the expansion were to continue while maintaining local equilibrium
so the generated density enhancements eventually subside.
However, as the hadronic gas grows more dilute, 
local equilibrium cannot be maintained. Consequently,
if the decoupling occurs sufficiently soon after the clumps are formed,
the associated phase-space correlations may survive. Such a correlation might be observed in a change of the width of two particle correlation functions \cite{Chen:2011vva}, even after the hadronic rescattering.  
This issue requires more detailed studies which are underway.

\begin{figure}[t]	
\includegraphics[width=0.5\textwidth]{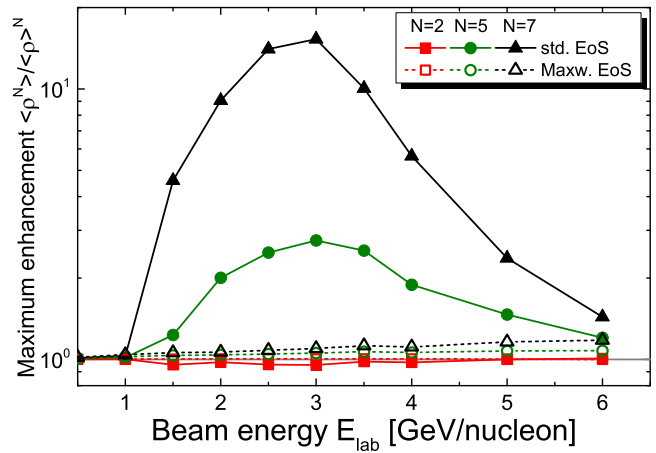}
\caption{[Color online] 
Mean maximum enhancement of the normalized density moments
for $N\!=\!7,5$ (squares, circles)
as obtained for various energies
using either the two-phase equation of state (solid)
or its one-phase Maxwell partner (dashed).
}\label{f:4}
\end{figure}		


The degree of density clumping generated during a collision
depends on how long time the bulk of the matter is exposed to
the spinodal instabilities.
The optimal situation occurs for collision energies that produce maximum
bulk compressions lying well inside the unstable phase region
because the instabilities may then act for the longest time
\cite{Randrup:2009gp,Randrup:2010ax,SteinheimerPRL109}.
At lower energies an ever smaller part of the system
reaches instability and the resulting enhancements are smaller.
Conversely, at higher energies the maximum compression occurs
beyond the spinodal phase region and the system is exposed to the
instabilities only during a relatively brief period
during the subsequent expansion.
For still higher energies the spinodal region is being missed entirely.

Figure \ref{f:4} shows the (ensemble average) maximum enhancement achieved
as a function of the beam energy for the two equations of state.
The existence of an optimal collision energy is clearly brought out.
While the presently employed \EoS\ suggests that this optimal range is
$E_{\rm lab}\approx2-4\,A\,\GeV$, it should be recognized that others
may lead to different results.
Furthermore, the magnitude of the effect depends on the degree of fluctuation
in the initial conditions which in turn are governed by the pre-equilibrium
dynamics.
On the other hand, our studies suggest that the optimal energy
is rather insensitive to the range parameter $a$.

\subsection{Size distribution}

\begin{figure}[t]	
\includegraphics[width=0.5\textwidth]{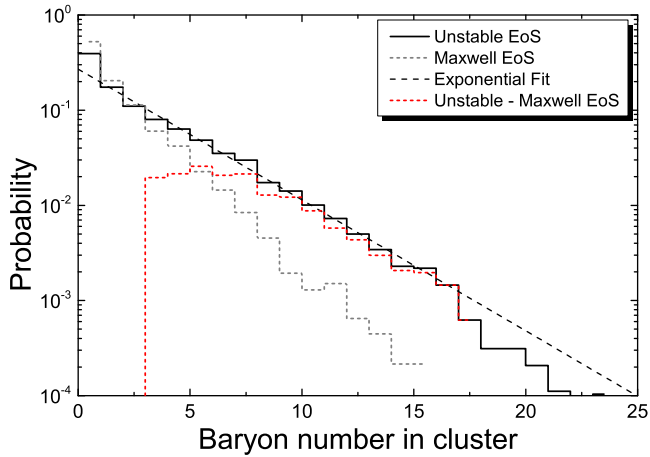}
\caption{\label{bnum} The size distribution of the density clumps
produced in central lead-lead collisions at $3~A\,\GeV$
using $\rho_{\rm min}=7 \rho_s$ to define the clump boundary.
The solid histogram shows the distribution obtained for the two-phase \EoS\
and the solid line represents an exponential fit.
The distribution obtained with the one-phase \EoS\ is shown by the
dotted histogram and the difference between the two size distributions
is also depicted.}
\end{figure}

To gain a more detailed understanding of the clumping phenomenon,
we have studied the distribution of the clump sizes.
Although the ``clumps'' tend to remain fairly diffuse,
we may define their extension by means of a specified density cutoff,
$\rho_{\rm min}$, and 
then extract the total net baryon number contained within the resulting volume.
Figure \ref{bnum} shows the size distribution obtained for a density cutoff of
$\rho_{\rm min}=7 \rho_s$,
for central lead-lead collisions at $3~A\,\GeV$.

The initial size distribution is approximately exponential
and that feature is well preserved during the evolution with the
one-phase \EoS\ which produces negligible amplification.
The spinodal instabilities in the two-phase \EoS\ leads to a
preferential amplification of length scales near the optimum size,
as is brought out by the difference between the two-phase size-distribution
and the one obtained with the Maxwell partner;
this difference peaks at clumps containing 5-8 baryons.
Nevertheless, for a wide intermediate range, from about 3 to about 16,
the resulting two-phase size distribution retains an approximately exponential
appearance, but with a significantly gentler slope.

\subsection{Transverse Flow}

Radial flow has long been thought to be a affected the presence of a
first-order phase transition because of the accompanying softening 
of the equation of state \cite{VanHove:1982vk}. 
It may therefore be expected that the transverse flow exhibits 
a corresponding weakening.
However, recent investigations using state-of-the-art dynamical models 
suggest that a simple connection between the softening of the \EoS\ 
and the radial flow (and its angular moments) 
cannot so easily be drawn \cite{Petersen:2009mz,Steinheimer:2009nn}. 
Although a softer \EoS\ does reduce the build up of flow ,
it also leads to a longer expansion time 
which tends to have an opposite effect.
 
We investigate here the effect of the phase transition on the transverse flow 
to determine  whether it may serve as a useful signal.
For this purpose, we define the transverse flow velocity $U_\perp$ as
\begin{equation}
U_\perp(t) = {1\over M(t)Q(t)}
	\int{[v_x(\bfr,t)\,x+v_y(\bfr,t)\,y]\,\rho(\bfr,t)\,d^3\bfr} ,
\end{equation}
where $\rho(\bfr)$ is the local (net) baryon density
at the position $\bfr=(x,y,z)$,
while $v_x(\bfr)$ and $v_y(\bfr)$ are the transverse components 
of the local fluid velocity $\boldsymbol{v}(\bfr)$.
 $M$ and $Q$ are defined as follows,
\begin{eqnarray}
	M(t)&=&\int{\rho(\bfr,t)\,d^3\bfr}\ ,\\
	M(t)Q(t)^2&=&\int{[x^2+y^2] \rho(\bfr,t)\,d^3\bfr}\ .
\end{eqnarray}

\begin{figure}[t]	
\center
\includegraphics[width=0.5\textwidth]{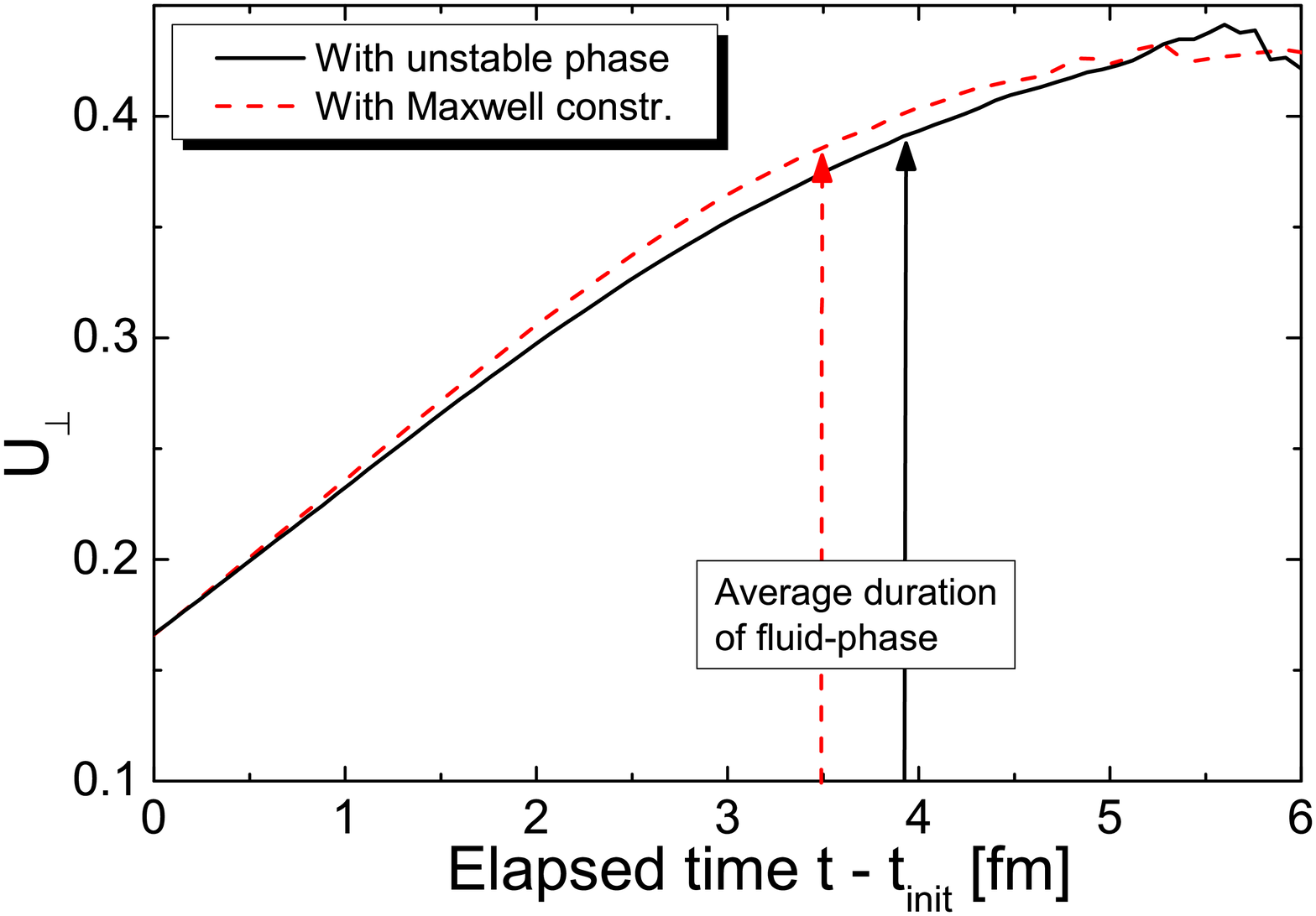}
\caption{\label{transv}(Color online) 
The time evolution of the average transverse flow velocity $U_\perp$
for central lead-lead collisions at a fixed-target 
beam kinetic energy of $E_{\rm lab}= 3\,A\,\GeV$
for the two-phase \EoS\ (solid) and its one-phase Maxwell partner (dashed).
The arrows indicate the average completion of the fluid-dynamical stage, 
defined as the time when all fluid cells have energy densities below 
five times the nuclear ground-state energy density.}
\end{figure}		

Figure \ref{transv} shows our results for the average transverse flow 
of baryons as a function of time for central lead-lead collisions
at the fixed-target beam energy of $E_{\rm lab}= 3\,A\,\GeV$
for which the density moments exhibited their largest enhancement.
Indeed transverse flow builds up more slowly for the two-phase \EoS\ 
than for its one-phase Maxwell partner,
though the difference between the two cases is very small,
probably too small to be of experimental utility.
Furthermore, the larger stiffness of the one-phase \EoS\
causes the expansion to be faster,
so the end of the the fluid-dynamical stage
(the time when the energy density has become sufficiently low,
$\eps=5 \eps_s$ in the present case) will occur sooner and this 
shorter expansion time will largely off-set the (small) gain in flow velocity.
As a result, the average transverse flow velocities
become almost identical at the end of the fluid-dynamical stage. 
Consequently, this quantity does not seem promising
as a signal for the phase transition.

\section{Parameter Studies}
The fluid dynamical model employed has several parameters
that are not well constrained by our current knowledge of QCD. 
These include the strength of the interface tension, 
the spatial scale of the initial density irregularities,
and the equation of state of baryon-rich uniform matter,
In the following,
we investigate the dependence of our results on these quantities 
by varying the respective controlling parameters one at a time.

\subsection{Finite range}
\begin{figure}[t]	
\center
\includegraphics[width=0.5\textwidth]{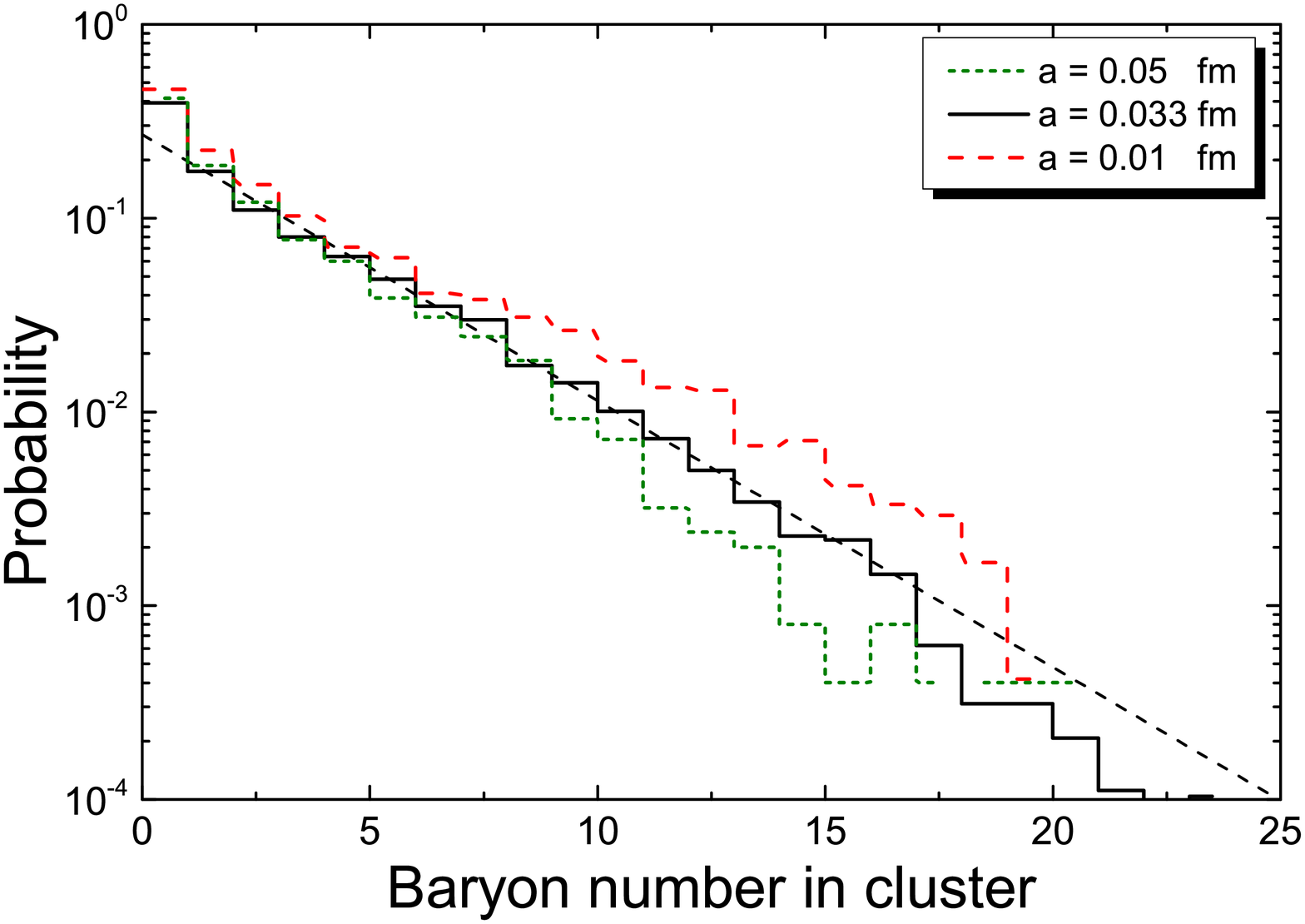}
\caption{\label{bnuma}(Color online) 
The size distribution of the density clumps
produced in central lead-lead collisions at $3~A\,\GeV$
as obtained for various values of the range parameter $a$
with the two-phase \EoS.
The dashed line shows an exponential fit to the standard result
(obtained for $a=0.033\,\fm$)}
\end{figure}		

\begin{figure}[t]	
\center
\includegraphics[width=0.5\textwidth]{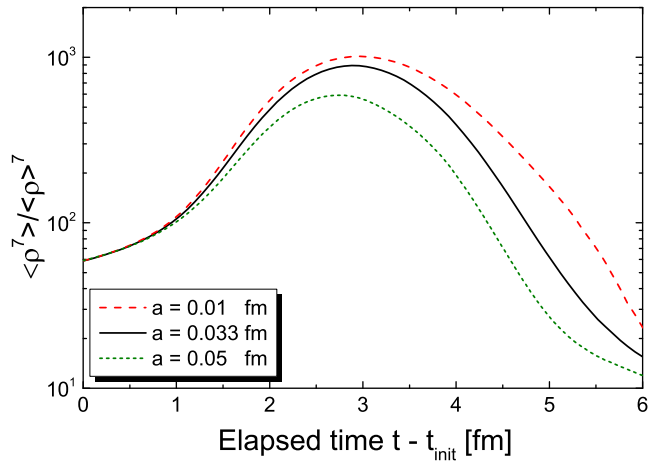}
\caption{\label{momta}(Color online) 
The time evolution of the seventh normalized density moment
for central lead-lead collisions at $E_{\rm lab}= 3A\,\GeV$
for three different values of the range parameter $a$,
using the standard value $\sigma_{\rm init}=1\,\fm$ for the spatial scale of the initial density irregularities.}
\end{figure}		

Figure \ref{bnuma} shows the dependence of the clump size distribution 
on the range parameter $a$ governing the interface tension
between baryon-rich confined and deconfined matter,
for the optimal collision energy.
We have shown above that density irregularities in the unstable phase region 
will grow faster for smaller ranges. 
This feature is also reflected in the size distribution:
a smaller $a$ value enhances the yield of larger clumps
({\em i.e.}\ density enhancements with larger baryon contents)
and, conversely,  a larger $a$ value decreases the large-$A$ yield.

Figure \ref{momta} shows how the variation in the range parameter $a$
affects the time dependence of the seventh normalized density moment. 
As expected, a smaller value of the range $a$ 
leads to a stronger enhancement of irregularities,
as already brought out in figure \ref{bnuma}.

\subsection{Initial fluctuations}

\begin{figure}[t]	
\center
\includegraphics[width=0.5\textwidth]{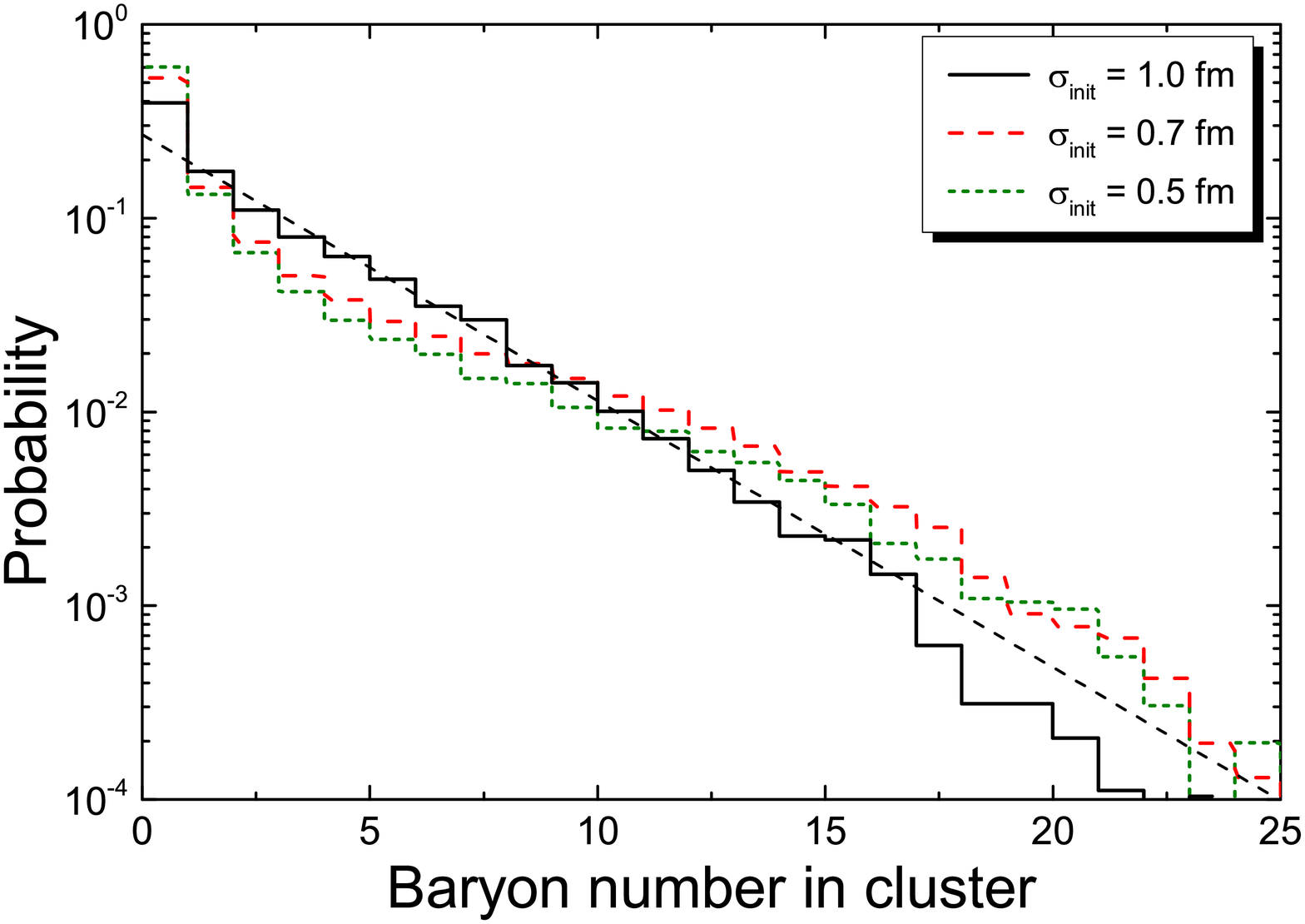}
\caption{\label{bnumsig}(Color online) 
The size distribution of the clumps
produced in central lead-lead collisions at $3~A\,\GeV$
obtained for different values of the parameter $\sigma_{\rm init}$
controlling the spatial scale of the initial irregularities,
using $a=0.033\,\fm$ and hte two-phase \EoS.
The dashed line shows an exponential fit to the standard result
(obtained for $\sigma_{\rm init}=1\,\fm$).
}
\end{figure}

Figure \ref{bnumsig} shows the dependence of the size distribution 
on the parameter $\sigma_{\rm init}$ 
characterizing the spatial scale of the initial density irregularities.
Although a reduction of the scale from the standard value of 
$\sigma_{\rm init}=1\,\fm$  shifts the clump sizes to larger values, 
the difference between the cases of $\sigma_{\rm init}=0.7 $ fm 
and $\sigma_{\rm init}=0.5 $ fm is very small,
indicating that the fluctuation growth saturates 
when the scale of the initial irregularities is reduced.

\begin{figure}[t]	
\center
\includegraphics[width=0.5\textwidth]{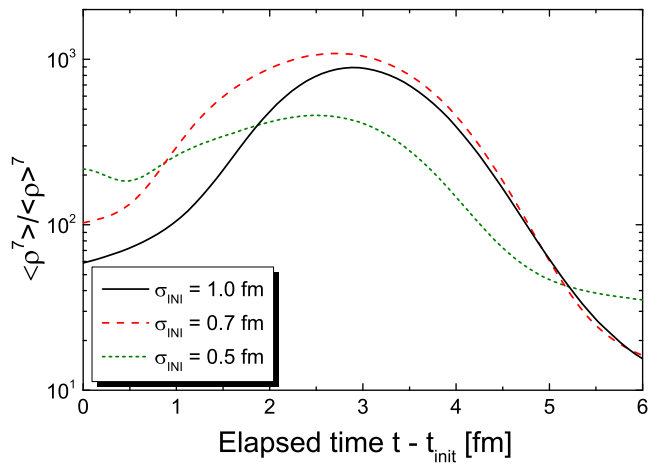}
\caption{\label{momtsig}(Color online) 
The time evolution of the seventh normalized moment
for central lead-lead collisions at $3A\,\GeV$
for three different values of the parameter $\sigma_{\rm init}$ 
controlling the scale of the initial density fluctuations,
using $a=0.033\,\fm$.}
\end{figure}

Figure \ref{momtsig} shows how the reduction of $\sigma_{\rm init}$
affects the evolution of the seventh normalized density moment. 
The behavior is somewhat complicated. 
While the initial value of the moment increases 
when the fluctuation scale $\sigma_{\rm init}$ is reduced,
its relative growth tends to become smaller. 
This feature may be ascribed to the fact that a system 
with larger initial density fluctuations also has larger density gradients,
which in turn implies that also a smaller part of the system 
lies inside the unstable phase region. 
Furthermore, the larger gradients also cause the expansion
through the unstable coexistence region to proceed more rapidly.

For our present studies, the initial state of the fluid-dynamical evolution
has been obtained by use of  the UrQMD model in its cascade mode
in which there are neither inter-nuclear potentials nor nuclear mean fields. 
However, for nuclear collisions at the relatively low energies
of relevance here, below $E_{\rm lab}\approx 5-10 A\,\GeV$, 
a realistic description of the nuclear interactions is important for
describing observables like the elliptic or directed flow 
\cite{Petersen:2006vm}. 
An UrQMD treatment of the initial state that includes mean-field type 
nuclear interactions \cite{Li:2006ez} 
will decrease the initial compression of the fireball. 

\begin{figure}[t]	
\center
\includegraphics[width=0.5\textwidth]{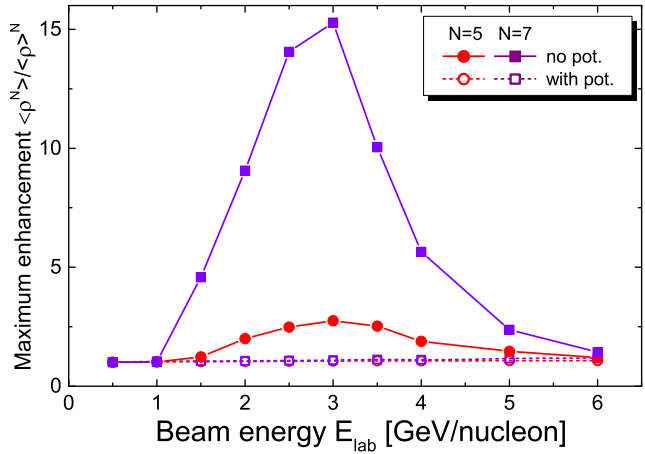}
\caption{\label{momelabini}(Color online) 
The collision energy dependence of the average maximum enhancement 
of the normalized density moments for $N =7, 5$  
as obtained for initial states generated
either by the cascade mode of the UrQMD model (our standard)
or with the inclusion of mean fields.}
\end{figure}

Figure \ref{momelabini} compares the maximum enhancements
of the fifth and seventh normalized density moments 
obtained with the cascade-mode initial conditions used in the present study
with those obtained when nuclear interactions are included. 
It is obvious that an enhancement is obtained only when the initial
density has been generated in the cascade mode.
This can be understood as a result of the considerably decreased 
initial compression due to the nuclear repulsive force 
which prevents the bulk of the system achieving the compressions
characteristic of the unstable phase region.
 
The results bring out the fact that 
the degree of density enhancement caused by the phase transition 
depend strongly on the initial compression 
relative to the location of the unstable region in the phase diagram.
The discussions on the mean-field	effect on the initial conditions are based on the hadronic mean fields in the UrQMD model,
which are strongly repulsive when the baryon density is large. Since the initial state in the present study is a baryon-rich
quark matter, the mean fields in such a matter may be different from those in hadronic matter. For example, the quark mean-field in the NJL	model is attractive due to the strong attractive scalar part (see Song et al. \cite{Song:2012cd} and references therein).

\subsection{Equation of State}

To further illuminate this latter feature,
we now modify the equation of state and investigate the dependence 
of the results on the location of the unstable region.

For this purpose, we consider a family of equations of state
obtained by rescaling the basic mechanical densities $\eps$ and $\rho$.
Thus, a particular scaled equation of state has the entropy density
$\sigma_{bc}(\eps,\rho) \doteq \sigma_0(b\eps,c\rho)$,
where $\sigma_0(\eps,\rho)$ is the entropy density of our standard \EoS.
It is then readily shown that the associated 
temperature and chemical potential are given by
\begin{equation}
T_{bc}(\eps,\rho)={1\over b}T_0(b\eps,c\rho)\ ,\,\
\mu_{bc}(\eps,\rho)={c\over b}\mu_0(b\eps,c\rho)\ ,
\end{equation}
where $T_0(\eps,\rho)=1/\del_\eps\sigma_0(\eps,\rho)$
is the temperature of the standard \EoS,
while the corresponding chemical potential is
$\mu_0(\eps,\rho)=-T_0(\eps,\rho)/\del_\rho\sigma_0(\eps,\rho)$.
It follows that the pressure $p=T\sigma-\eps+\mu\rho$ is given by
\begin{equation}
p_{bc}(\eps,\rho)={1\over b}p_0(b\eps,c\rho)\ .
\end{equation}
The enthalpy and the free energy scale similarly, {\em i.e.}\
$h_{bc}(\eps,\rho)=h_0(b\eps,c\rho)/b$ and
$f_{bc}(\eps,\rho)=f_0(b\eps,c\rho)/b$.
It is also easy to see that the compressibilities remain invariant,
$c_v^{bc}(\eps,\rho)=c_v^0(b\eps,c\rho)$ and
$c_p^{bc}(\eps,\rho)=c_p^0(b\eps,c\rho)$.
The same holds for the sound speeds,
\begin{equation}
v_s^{bc}(\eps,\rho)=v_s^0(b\eps,c\rho)\ ,\,\
v_T^{bc}(\eps,\rho)=v_T^0(b\eps,c\rho)\ ,
\end{equation}
thus ensuring that the isentropic and isothermal spinodal boundaries
stretch in proportion to the employed scale factors.
In particular, the scaled critical point is given by
$(\eps_{bc}^{\rm crit},\rho_{bc}^{\rm crit})
=(\eps_0^{\rm crit}/b,\rho_0^{\rm crit}/c)$.\\

\begin{figure}[t]	
\center
\includegraphics[width=0.5\textwidth]{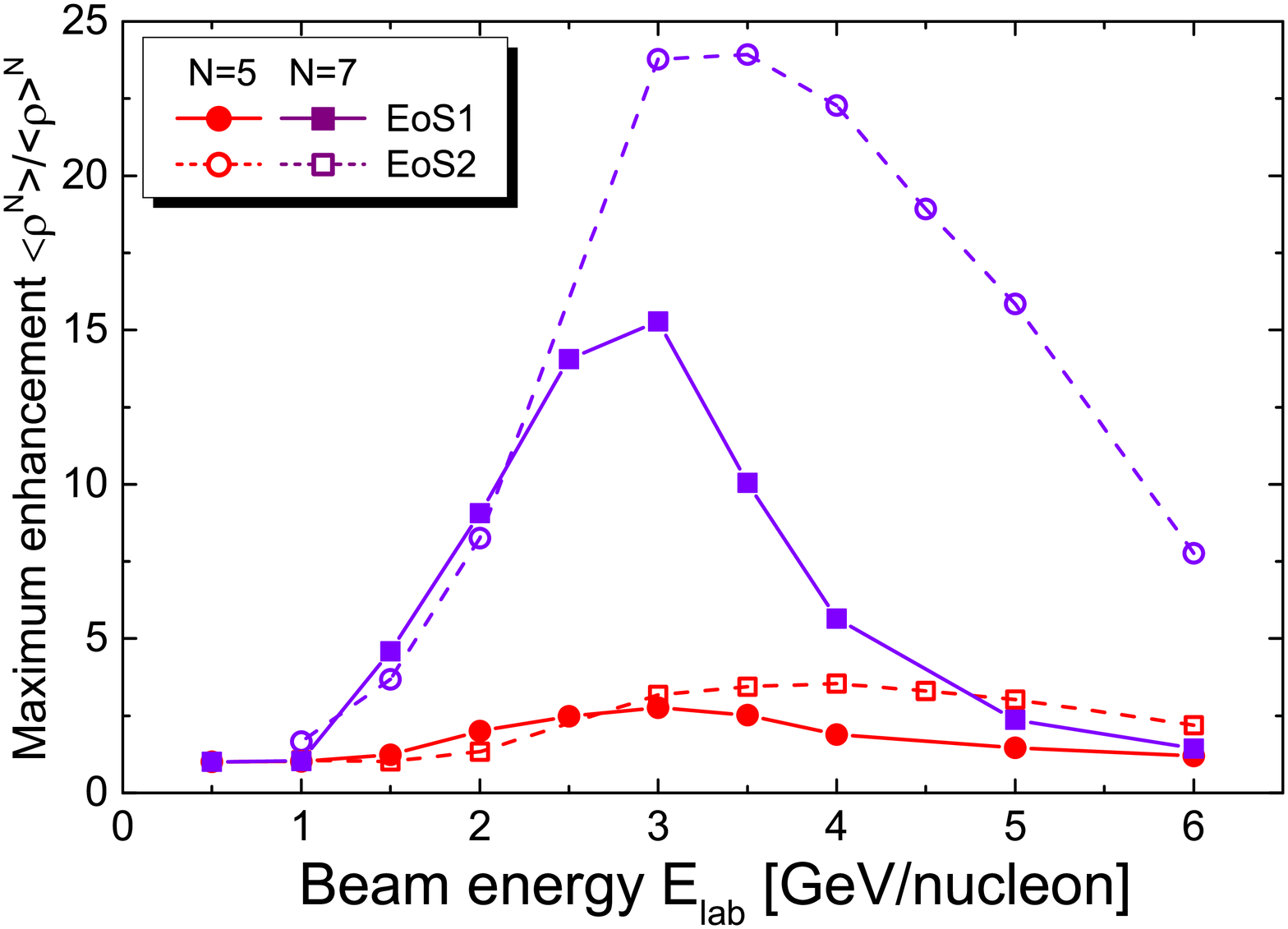}
\caption{\label{momelabeos}(Color online)
The average maximum enhancement of the normalized density moments 
for $N=7,5$ as a function of the beam kinetic energy $E_{\rm lab}$,
obtained with either our standard \EoS\ (solid curves)
or the scaled one corresponding to $b=1/1.3$ and $c=1/1.3$ (dashed curves).}
\end{figure}

The scaling properties exhibited above provides a convenient means
for exploring different equations of state.
As an illustration, Fig.\ \ref{momelabeos} compares the energy dependence 
of the density enhancements obtained with our standard \EoS\ 
to those obtained with an \EoS\ that has been stretched upwards
by 30\%\ in both directions, {\em i.e.}\ $b=1/1.3$ and $c=1/1.3$.
We see that the qualitative features remain very similar.
In particular, there is still an optimal collision energy.
The larger enhancement obtained for the scaled \EoS\ is primarily
due to the corresponding expansion of the unstable phase region
which allows the amplification to proceed for a longer time.
However, the optimal collision energy is only slightly increased.

\section{Summary}

As reported recently \cite{SteinheimerPRL109},
we have augmented an existing finite-density ideal fluid dynamics code
with a gradient term and thereby obtained a transport model 
that is suitable for simulating nuclear collisions 
in the presence of a first-order phase transition.
It describes both the temperature-dependent tension between coexisting phases
and the amplification of the spinodal modes.
Applying this novel model to lead-lead collisions,
using an \EoS\ with a first-order phase transition,
we found that the associated instabilities may cause
significant amplification of initial density irregularities,
relative to what would be obtained without the phase transition.
Because such clumping may give rise to observable effects
that could signal the presence of the phase transition,
perhaps through enhanced composite-particle production,
it is interesting to explore this phenomenon in more detail.

We have here given a more detailed account of this novel simulation tool
and carried out more exhaustive studies of the phase-separation dynamics.
In particular, we extracted the density enhancement,
the clump size distribution, and the transverse flow velocity
and examined the sensitivity of these quantities 
to the strength of the gradient term that promotes the phase separation,
to the details of the initial density fluctuations
that form the seeds for the subsequent amplification,
and to the equation of state,
all quantities that are yet not very well understood.
We found that our results are robust 
against moderate changes of these model parameters. 

While the presence of a phase transition greatly increases
the density enhancements,
it has only little effect on the resulting transverse flow.
Perhaps most importantly, studies using different equations of state
support the general existence of an optimal collision energy range
within which the phase-transition instabilities have the largest
effects on the dynamical evolution.
Our results suggest that this energy corresponds to several GeV per nucleon
of kinetic energy for a fixed-target configuration,
a range that may be too low to access effectively at RHIC
but which should match well with both FAIR and, especially, NICA.

\section{Acknowledgments}
We wish to acknowledge stimulating discussion with Volker Koch.
This work was supported by the Office of Nuclear Physics
in the U.S.\ Department of Energy's Office of Science
under Contract No.\ DE-AC02-05CH11231;
JS was supported in part by the Alexander von Humboldt Foundation
as a Feodor Lynen Fellow.


\end{document}